\documentclass[%
 reprint,
 amsmath,amssymb,
 aps,
 pre,
 showkeys,
 floatfix,
]{revtex4-2}

\usepackage{graphicx}% Include figure files
\usepackage{dcolumn}% Align table columns on decimal point
\usepackage{bm}% bold math
\usepackage[hidelinks]{hyperref}
\begin{document}

\title{Ensemble-level loopy message passing with generalized-edge closure for percolation}

\author{L.-H. Wang}
\author{Y.-M. Du}
\email{ymdu@gscaep.ac.cn}
\affiliation{Graduate School of China Academy of Engineering Physics, Beijing 100193, China}

\date{\today}

\begin{abstract}
Predicting the percolation threshold of highly clustered networks from local statistics remains difficult, because short loops break the independence assumption underlying tree-like message passing. Existing remedies address loopy connectivity either through prescribed local motifs in random-graph ensembles or through a single network's realized topology, leaving an ensemble-level treatment of arbitrary connectivity patterns absent. Here, we develop a loopy message-passing framework for random clustered graph ensembles based on generalized-edge statistics, which characterize overlap patterns among the neighborhoods of different nodes. This yields a progressively refined approximation scheme based on neighborhoods of increasing size around each node. The low-order approximations recover previous equations for random network ensembles, and the new result that yields refined threshold prediction is developed by the second-order approximation. We show that the effectiveness of this framework depends not only on short-cycle density but also on the internal consistency of generalized edges. To diagnose this effectiveness, we introduce the generalized-edge closure coefficient (GECC) to quantify this consistency. Because GECC is computed entirely from local statistics and does not rely on any percolation calculation, it serves as an a priori diagnostic for the reliability of the approximation. Using synthetic and real networks, the threshold is evaluated via the second-order and lower-order approximations. Comparisons with Monte Carlo simulations show that GECC captures key structural features that strongly affect the percolation threshold. These results establish ensemble-based loopy message passing as an efficient route for predicting the percolation threshold in large clustered networks.
\end{abstract}


\maketitle

%\tableofcontents

\section{Introduction}

Percolation provides a fundamental framework for characterizing network
robustness, describing when large-scale connectivity emerges or collapses in
complex networks
\cite{Newman2003,Callaway2000,Cohen2000,Cohen2001,Albert2000,MooreNewman2000,ArtimeDeDomenico2021,Wlh}.
The percolation threshold links local network structure to macroscopic
connectivity transitions and supports analyses of resilience, epidemics, and
influence propagation
\cite{Karrer2010,Karrer2014,Morone2015,Qian2024}.
Message-passing (MP) methods are a standard analytical tool for such problems,
propagating local connectivity information along edges
\cite{Karrer2010,Karrer2014,Qian2024}; related belief-propagation (BP) methods are
widely used in probabilistic inference and statistical physics~\cite{Pearl1988,Yedidia2005,Mezard2009}.
Their accuracy typically relies on the locally tree-like approximation, which
treats incoming messages as independent~\cite{Newman2001,Dorogovtsev2008,PastorSatorrasVespignani2001,KenahRobins2007,Karrer2014}.
In highly clustered networks, short loops such as triangles and four-cycles
violate this independence assumption, producing local correlations and
systematic errors in connectivity and threshold predictions~\cite{Watts1998,Newman2009,Miller2009}. This breakdown reflects a generic failure of the independence assumption underlying MP/BP, rather than a feature specific to percolation, and similarly affects MP-based calculations of epidemic spreading and probabilistic inference on clustered networks~\cite{Karrer2010,Pearl1988}. Reliable threshold prediction in such networks therefore requires MP
frameworks that explicitly account for loopy local connectivity patterns~\cite{Barrat2008,Newman2010,CohenHavlin2010,RadicchiCastellano2015,RadicchiBianconi2017}.

Early efforts to move beyond the locally tree-like approximation incorporated triangle-type clustering into random graph ensembles.
Such clustering was characterized either through degree-dependent statistics of triangular closure~\cite{Serrano2006,Serrano2006b,Serrano2006PRL} or by treating triangles as local building blocks embedded in an effectively tree-like backbone~\cite{Newman2009,Miller2009,Miller2009JRSI}. This building-block perspective was later extended from triangles to cliques and motifs within the same tree-like organization~\cite{Newman2003Clustered,Britton2008,Gleeson2009,BollobasJansonRiordan2011,KarrerNewman2010,Allard2012}.
Percolation on these graph ensembles is tractable because they are equipped with prior information about both the tree-like organization among building blocks and the distribution of the building blocks themselves.
Theories for arbitrary local connectivity patterns have also been developed for deterministic networks~\cite{MooijKappen2007,Cantwell2019,MannDobson2023}.
These approaches naturally form a progressively refined approximation scheme.
In the formulation of Ref.~\cite{Cantwell2019}, the lowest-order approximation recovers tree-like message passing, whereas higher orders compute messages by excluding increasingly larger neighborhoods around the target node, thereby reducing feedback from short loops and improving the treatment of loop-induced correlations.
Together, these lines of work address loopy local connectivity either through prescribed local building blocks in random graph ensembles or through the realized topology of a deterministic network.

The remaining gap is to construct a message-passing framework for random graph ensembles with arbitrary local connectivity patterns using only local statistics.
The central difficulty is to propagate messages without access to a fixed realized topology: each transmission depends not only on the neighborhood being entered but also on how that neighborhood is embedded in its surroundings.
If this embedding information is not retained during propagation, correlations generated by short loops are lost.
The key task is therefore to identify, from local statistics, the structural features that must be retained to close the message-passing equations in the presence of loop-induced correlations, without assuming either prescribed local building blocks or a realized topology.

Guided by the progressively refined perspective of Ref.~\cite{Cantwell2019}, we develop a loopy message-passing framework for random clustered graph ensembles.
To our knowledge, this is the first ensemble-level formulation of such a progressively refined neighborhood-based approach for random-graph ensembles.
At the ensemble level, messages are associated with jointly sampled neighborhood pairs rather than with fixed edges of a realized graph.
The overlaps of these neighborhoods define generalized edges, whose transition statistics retain the embedding information needed to propagate loop-induced correlations.
We derive nonlinear ensemble-level recursions and explicitly work out the second-order approximation, obtaining tractable message-passing equations through linearization and feature-conditioned reduction.
The internal consistency of generalized edges determines the accuracy of the approximation; we call this property generalized-edge closure and quantify it by the generalized-edge closure coefficient, \ensuremath{\mathrm{GECC}}.
Tests on synthetic and real networks show that the resulting theory captures loop-induced correlations, identifies regimes where tree-like or first-order loopy approximations break down, and improves estimates of percolation thresholds in highly clustered networks. Because GECC is computed entirely from local ensemble statistics and is independent of any specific dynamical process, it provides an a priori diagnostic for the reliability of loopy message-passing approximations that is, in principle, applicable beyond percolation to other MP-based calculations on clustered networks.
Together, these results show how closure-relevant structural features can be extracted from ensemble statistics, establishing a tractable and effective framework for loopy message passing in highly clustered random graph ensembles.

\section{Results}\label{sec2}

\subsection{Order-dependent subgraphs in random graphs}

We adapt the notation of Ref.~\cite{Cantwell2019} to random graphs. In a fixed network, the $n$th-order
subgraph $\Gamma_i^{(n)}$ around node $i$ retains paths of length at most $n$
between neighbors of $i$, and therefore resolves closed paths involving $i$
up to length $n+2$. For random graphs, we replace the instance-specific
subgraph $\Gamma_i^{(n)}$ by a degree-conditioned local type
$\Gamma_k^{(n)}$, which describes the ensemble-level local connectivity patterns around
nodes of degree $k$. We define $N_1(k)$ as the set of shell-1 neighbors of a degree-$k$ node, i.e.,
its direct neighbors. For $r\geq 2$, let $N_r(k)$ be the set of nodes that belong to $\Gamma_k^{(r)}$ but not to $\Gamma_k^{(r-1)}$.
Nodes in $N_r(k)$ are called the shell-$r$ neighbors of the degree-$k$ node. A detailed definition is provided in the Supplemental Material.

\subsection{Loopy message-passing through Generalized-edge}
\subsubsection{Generalized-edge messages}

The progressively refined formulation replaces single-edge transmission with
message passing on generalized edges. For subgraphs of $n$th-order, consider two
subgraphs $\Gamma_k^{(n)}$ and $\Gamma_d^{(n)}$ associated with two nodes of
degrees $k$ and $d$, respectively. The generalized edge formed by these two
nodes is defined as
\begin{equation}
E_{kd}^{(n)}=\Gamma_k^{(n)}\cap \Gamma_d^{(n)},
\end{equation}
and is referred to as an $n$th-order generalized edge between these two nodes.
The ordinary edge corresponds to the lowest-order generalized edge $n=0$, for
which $E_{kd}^{(0)}$ contains only the direct connection between the two nodes.

In the tree-like formulation, messages are transmitted along ordinary edges,
while incoming-message collection is determined by the degree of the node
receiving the message. This degree can be interpreted as the subgraph
at the lowest order. In higher-order approximations, subgraphs are enlarged.
Consider a message collected at a node characterized by $\Gamma_d^{(n)}$ and
then transmitted to a node characterized by $\Gamma_k^{(n)}$. The generalized
edge supporting this transmission is $E_{kd}^{(n)}$. The subgraph
$\Gamma_d^{(n)}$ can therefore be separated into $E_{kd}^{(n)}$ and the
residual structure
\begin{equation}
R_{d\to k}^{(n)}=\Gamma_d^{(n)}\setminus E_{kd}^{(n)},
\end{equation}
which is used for collecting incoming messages. Thus, although the generalized
edge itself is specified by $E_{kd}^{(n)}$, the message must be specified by
the subgraph pair $(\Gamma_k^{(n)},\Gamma_d^{(n)})$. Let
$\pi(s\mid\Gamma_k^{(n)},\Gamma_d^{(n)})$ be the probability that a
degree-$k$ focal node reaches a finite cluster of size $s$ through a
degree-$d$ node in its $n$th-order subgraph. We decompose this message as
\begin{equation}\label{2}
\begin{split}
\pi(s \mid \Gamma_k^{(n)},\Gamma_d^{(n)})
={}&
\tilde{\pi}(s \mid \Gamma_k^{(n)},\Gamma_d^{(n)})
g(q,\Gamma_k^{(n)}\cap\Gamma_d^{(n)})  \\
&+
\delta_{s,0}
\left[
1-g(q,\Gamma_k^{(n)}\cap\Gamma_d^{(n)})
\right].
\end{split}
\end{equation}
Here $g(q,\Gamma_k^{(n)}\cap\Gamma_d^{(n)})$ is the probability that, under
occupation probability $q$, the two subgraphs $\Gamma_k^{(n)}$ and
$\Gamma_d^{(n)}$ are connected through their generalized edge by at least one
occupied path. Conditional on this connection event,
$\tilde{\pi}(s\mid\Gamma_k^{(n)},\Gamma_d^{(n)})$ is the probability that the
degree-$k$ focal node reaches a finite cluster of size $s$ through the
degree-$d$ node. Thus, each subgraph pair contributes two quantities
to message passing: the connection probability
$g(q,\Gamma_k^{(n)}\cap\Gamma_d^{(n)})$ and the conditional
finite-cluster-size distribution
$\tilde{\pi}(s\mid\Gamma_k^{(n)},\Gamma_d^{(n)})$. This mirrors the split of
$\Gamma_d^{(n)}$ into $E_{kd}^{(n)}$ and $R_{d\to k}^{(n)}$:
$g(q,E_{kd}^{(n)})$ depends only on $E_{kd}^{(n)}$, while
$\tilde{\pi}(s\mid\Gamma_k^{(n)},\Gamma_d^{(n)})$ depends on
$R_{d\to k}^{(n)}$.

\subsubsection{Loopy message-passing formulation on random networks}
With the generalized edges and the associated probability quantities defined
above, we now derive the central technical result of this framework: an exact, ensemble-level message-passing recursion, Eq.~(\ref{3}), that holds
for arbitrary random networks at arbitrary approximation order $n$. All correlations induced by overlapping generalized edges are absorbed into a single structural function, $\mathcal{A}^{(n)}$, which we show below factorizes using the independence of messages from shell-1 neighbors.

\begin{figure*}[t]
\begin{equation}\label{3}
\begin{aligned}
&
\tilde{\pi}\!\left(s\mid \Gamma_k^{(n)},\Gamma_d^{(n)}\right)
g\!\left(q,\Gamma_k^{(n)}\cap\Gamma_d^{(n)}\right)
+
\delta_{s,0}
\left[
1-g\!\left(q,\Gamma_k^{(n)}\cap\Gamma_d^{(n)}\right)
\right]
\\
&\quad =
\delta_{s,0}
\left[
1-g\!\left(q,\Gamma_k^{(n)}\cap\Gamma_d^{(n)}\right)
\right]
+
q\,g\!\left(q,\Gamma_k^{(n)}\cap\Gamma_d^{(n)}\right)
\sum_{\{\Gamma_{l_i}^{(n)}\}_{i=1}^{m}}
\sum_{s_1,\dots,s_m}
\mathcal{A}^{(n)}
\left(
s_1,\dots,s_m
\,\Bigm|\,
\Gamma_d^{(n)},
\Gamma_{l_1}^{(n)},\dots,\Gamma_{l_m}^{(n)}
\right)
\\
&\qquad \times
\delta
\left(
\sum_{i=1}^{m}s_i-s+1
\right)
\Pr
\left(
\Gamma_{l_1}^{(n)},\dots,\Gamma_{l_m}^{(n)}
\,\Bigm|\,
\Gamma_k^{(n)},\Gamma_d^{(n)}
\right).
\end{aligned}
\end{equation}
\end{figure*}

\begin{figure*}[t]
    \centering
    \includegraphics[width=\textwidth]{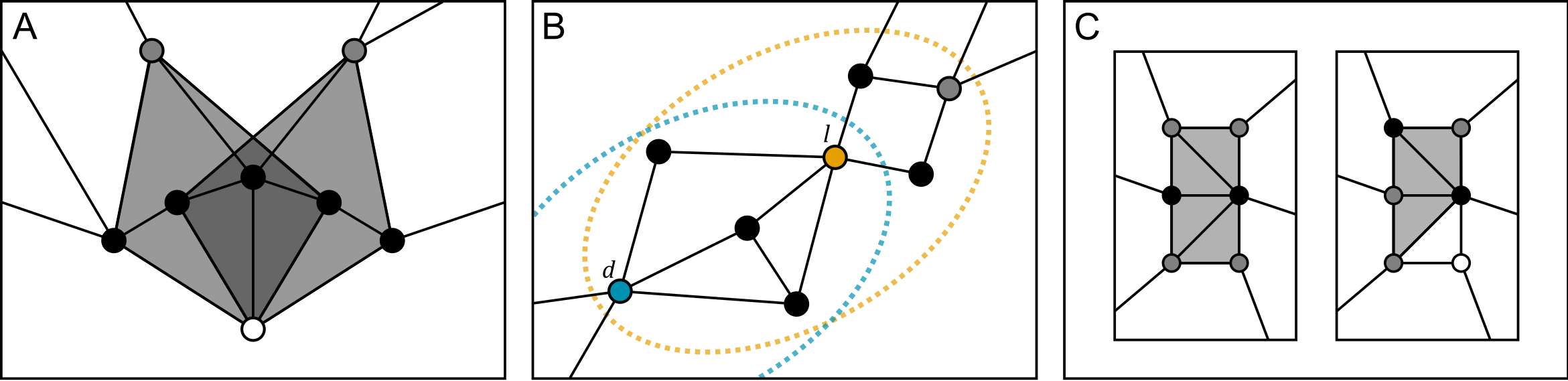}
    \caption{Schematic illustration of second-order generalized edges, their closure
property, and associated structural features.
\textbf{(A)} Correlation structure encoded in \ensuremath{\mathcal{A}^{(2)}} when the
number of shell-2 neighbors is two. The white node represents the focal node
of degree \ensuremath{d}, black nodes represent its shell-1 neighbors in
\ensuremath{N_1(d)}, and gray nodes represent its shell-2 neighbors in
\ensuremath{N_2(d)}. Gray shaded regions denote generalized edges between the
focal node and its shell-2 neighbors, and the darker gray region indicates the
overlap between different generalized edges.
\textbf{(B)} Structural features of a second-order generalized edge. The blue node and
dashed curve indicate the node of degree \ensuremath{d} and its
second-order subgraph \ensuremath{\Gamma_d^{(2)}}, whereas the orange
node and dashed curve indicate the target node of degree \ensuremath{l}
and its second-order subgraph \ensuremath{\Gamma_l^{(2)}}. In this
example, the two nodes are separated by graph distance \ensuremath{y=2}
within the generalized edge, with \ensuremath{w=3} shared shell-1
neighbors, and the residual part of the target node subgraph gives
$m=\left|\Gamma_l^{(n)}\setminus(\Gamma_d^{(n)}\cap\Gamma_l^{(n)})\right|=3$.
\textbf{(C)} Generalized-edge closure. The left subpanel shows the initial generalized
edge determined by a selected pair of black nodes. The right subpanel shows
the generalized edge induced by another pair of black nodes selected within
the initial generalized edge. The induced generalized edge differs from the
initial one, indicating poor closure of the initial generalized edge under
internal node-pair selections. Gray nodes and gray shaded regions indicate
nodes and internal edges contained in the corresponding generalized edges,
whereas white nodes indicate nodes outside the current generalized edge.}
    \label{fig:ge_schematic_summary}
\end{figure*}

In Eq.~(\ref{3}), $m=\left|\Gamma_d^{(n)}\setminus\left(\Gamma_k^{(n)}\cap\Gamma_d^{(n)}\right)\right|$
counts the residual nodes in $\Gamma_d^{(n)}$ outside the generalized edge
$E_{kd}^{(n)}$ (for any graph $\Gamma$, $|\Gamma|$ is its node count); each
residual node $i=1,\dots,m$ carries a cluster-size contribution $s_i$ and an
$n$th-order subgraph $\Gamma_{l_i}^{(n)}$ determined by its degree $l_i$. The
Kronecker delta enforces the size constraint $\sum_i s_i=s-1$, and
$\Pr(\Gamma_{l_1}^{(n)},\dots,\Gamma_{l_m}^{(n)}\mid\Gamma_k^{(n)},\Gamma_d^{(n)})$
gives the joint probability of these residual subgraphs conditional on the
message-defining pair $(\Gamma_k^{(n)},\Gamma_d^{(n)})$.
$\mathcal{A}^{(n)}$ concentrates the joint correlation between the inner
shell-1 neighbors and the outer shell-$r$ ($r\geq2$) neighbors, thereby
replacing the usual product of independent external messages used in the
tree-like approximation. Formally, $\mathcal{A}^{(n)}$ is defined as the
conditional joint distribution of the finite-cluster sizes
$\{s_i\}_{i=1}^{m}$ contributed by the residual nodes, given $\Gamma_d^{(n)}$
and the associated subgraphs $\Gamma_{l_1}^{(n)},\dots,\Gamma_{l_m}^{(n)}$.
Although these correlations become more involved as $n$ increases, the
message contributions from the nodes in $N_1(d)$ remain mutually
independent because they are direct neighbors of the degree-$d$ node.
Therefore, $\mathcal{A}^{(n)}$ can be factorized into a conditional
probability for the shell-$r$ neighbors with $r\geq 2$ and independent
shell-1-neighbor message contributions:
\begin{equation}\label{3+}
\begin{aligned}
&\mathcal{A}^{(n)}
\left(
s_{1},\dots,s_{m}
\,\Bigm|\,
\Gamma_d^{(n)},
\Gamma_{l_1}^{(n)},\dots,\Gamma_{l_m}^{(n)}
\right)
\\
=&\,
\Pr\Bigg(
\{s_j\}_{j\in \bigcup_{r=2}^{n}N_r(d)}
\,\Bigm|\,
\{s_i\}_{i\in N_1(d)},\Gamma_d^{(n)},\Gamma_{l_1}^{(n)}
\\
&\qquad
,\dots,\Gamma_{l_m}^{(n)}
\Bigg)
\prod_{i\in N_1(d)}
\tilde{\pi}\left(
s_i
\,\middle|\,
\Gamma_d^{(n)},\Gamma_{l_i}^{(n)}
\right).
\end{aligned}
\end{equation}
When $n=2$, Eq.~(\ref{3+}) has a clear structural meaning: the conditional
probability gives the joint contribution of the shell-2 neighbors $N_2(d)$,
conditioned on the shell-1 contributions, since each shell-2 neighbor reaches
the degree-$d$ node only through a shell-1 neighbor. Two shell-2 neighbors
sharing the same shell-1 neighbor therefore depend on the same shell-1
message, corresponding to an overlap between their generalized edges that
defines the correlation structure of $\mathcal{A}^{(2)}$, as illustrated in
Fig.~\ref{fig:ge_schematic_summary}(A).

Although Eq.~(\ref{3}) is high-dimensional, determining the percolation
threshold only requires its linear stability analysis, which reduces the
problem to the tractable self-consistent recursion, Eq.~(\ref{5}). For the
$n$th-order approximation ($n=0,1,2$), we define
$\epsilon_{\Gamma_k^{(n)},\Gamma_d^{(n)}}$ as the probability that a
degree-$k$ focal node reaches the giant component through a degree-$d$ node
in its $n$th-order subgraph, given by
$\epsilon_{\Gamma_k^{(n)},\Gamma_d^{(n)}} = 1-H(1\mid\Gamma_k^{(n)},\Gamma_d^{(n)})$,
where $H(z\mid\Gamma_k^{(n)},\Gamma_d^{(n)})$ denotes the generating
function of $\pi(s\mid\Gamma_k^{(n)},\Gamma_d^{(n)})$ with parameter $z$.
The resulting message-passing equation is
\begin{equation}\label{5}
\begin{aligned}
\epsilon_{\Gamma_k^{(n)},\Gamma_d^{(n)}}=&q\sum_{\Gamma_{l}^{(n)}} \left|\Gamma_d^{(n)}\setminus(\Gamma_k^{(n)}\cap\Gamma_d^{(n)})\right|g(q,\Gamma_d^{(n)}\cap\Gamma_l^{(n)})\\
&\epsilon_{ \Gamma_d^{(n)},\Gamma_l^{(n)}} \Pr\left(\Gamma_{l}^{(n)}|\Gamma_k^{(n)},\Gamma_d^{(n)}\right).
\end{aligned}
\end{equation}
 The correlation structure encoded in $\mathcal{A}^{(2)}$ is
detailed in Supplemental Material, and the derivation of Eq.~(\ref{5}) is also
provided there.

Although the resulting equations are formulated at the ensemble level, they are
still not directly tractable. We therefore extract typical structural features
from local statistics to construct a reduced self-consistent message-passing equation.

\subsection{Message passing based on typical structural features}

\subsubsection{Extraction of typical structural features}

For an $n$th-order generalized edge associated with subgraphs
$\Gamma_d^{(n)}$ and $\Gamma_l^{(n)}$, we extract a minimal set of
structural features needed by the reduced message-passing equation: the
degree classes involved in the transition; the effective connectivity,
extracted from the generalized edge $E_{dl}^{(n)}$; and the size available
for subsequent propagation, extracted from the residual structure
$R_{l\to d}^{(n)}$. This reduction yields a tractable message-passing
formulation by encoding generalized edges and subsequent-propagation
components in terms of a finite set of structural features.

\subsubsection{Second-order message passing equation}

For a second-order generalized edge associated with subgraphs
$\Gamma_d^{(2)}$ and $\Gamma_l^{(2)}$, the corresponding message-passing
equation retains three types of structural features: the node degrees $d$
and $l$; the effective connectivity within the generalized edge
$E_{dl}^{(2)}$, specified by the distance $y$ and the number $w$ of shared
shell-1 neighbors; and the size of the residual structure
$R_{l\to d}^{(2)}$ available for subsequent propagation, quantified by the
residual size $m$. An example is shown in
Fig.~\ref{fig:ge_schematic_summary}(B).

The graph-dependent variables in Eq.~(\ref{5}) can then be expressed as
functions of the second-order structural features:
\begin{equation}\label{6+}
\begin{aligned}
\epsilon_{\Gamma_d^{(2)},\Gamma_l^{(2)}}
&\longrightarrow
\epsilon_l,\\
\Pr\left(\Gamma_l^{(2)}
\mid \Gamma_k^{(2)},\Gamma_d^{(2)}\right)
&\longrightarrow
P(l,y,w\mid d).
\end{aligned}
\end{equation}
Here, \ensuremath{\epsilon_l} denotes the probability that an edge pointing
to a node of degree \ensuremath{l} reaches the giant component. The term
\ensuremath{P(l,y,w\mid d)} denotes the probability that, in the
second-order subgraph of a node with degree \ensuremath{d}, a node has
degree \ensuremath{l}, distance \ensuremath{y} from the degree-\ensuremath{d}
node, and \ensuremath{w} shared shell-1 neighbors. The graph-dependent connection factor $g(q,\Gamma_d^{(2)}\cap\Gamma_l^{(2)})$
is expressed as $g(q,y,w)=1-(1-q^{y-1})^w$, which gives the probability that
connectivity is established through at least one occupied shared shell-1
neighbor.

For analytical tractability, we replace $m$ by its average over generalized
edges with the same values of $d,l,y$ and $w$. Let
$P(m\mid d,l,y,w)$ denote the probability that, among second-order
generalized edges with degrees $d,l$, distance $y$, and $w$ shared
shell-1 neighbors, the residual structure $R_{l\to d}^{(2)}$ has size $m$.
The resulting average residual size is
\begin{equation}\label{6++}
m_{dlyw}^{(2)}=\sum_{m} m\, P(m\mid d,l,y,w).
\end{equation}

With these reductions, relabeling the current degree class \ensuremath{d}
as \ensuremath{k} and the next degree class \ensuremath{l} as
\ensuremath{d} gives
\begin{equation}\label{9}
\epsilon_k
=
q
\sum_{d,y,w}
m_{kdyw}^{(2)}
g(q,y,w)
\epsilon_d
P(d,y,w\mid k).
\end{equation}
Eq.~(\ref{9}) is obtained from Eq.~(\ref{3}) through standard linearization and structural reduction. The resulting structural template already
captures the full form of the generalized-edge framework: higher orders $n\geq2$ retain the same combination of degree, distance-type
connectivity, and residual-size variables, differing from Eq.~(\ref{9})
only in computational complexity rather than in structural nature. Eq.~(\ref{9})
therefore constitutes the first nontrivial tractable realization of the progressively refined ensemble framework for arbitrary order $n$.

\subsubsection{Recovery of lower-order approximations}

We next show how lower-order approximations are recovered from
lower-order subgraphs. At both first and zeroth orders, only shell-1
neighbors are retained, and hence the second-order connectivity
variables $y$ and $w$ do not enter the reduced description. As a
result, the corresponding connection factor is set to
$g(q,y,w)=1$. For $n=0$ or $n=1$, the graph-dependent variables are
then reduced to
\begin{equation}\label{6_lower}
\begin{aligned}
\epsilon_{\Gamma_d^{(n)},\Gamma_l^{(n)}}
&\longrightarrow
\epsilon_l,\\
\Pr\left(\Gamma_l^{(n)}
\mid \Gamma_k^{(n)},\Gamma_d^{(n)}\right)
&\longrightarrow
P(l\mid d).
\end{aligned}
\end{equation}
Here, \ensuremath{\epsilon_l} is defined as above, and $P(l\mid d)$
denotes the probability that an edge of a degree-$d$ node points to
a degree-$l$ node. After relabelling, they become
\ensuremath{\epsilon_d} and \ensuremath{P(d\mid k)} below.

At first order, the subgraph contains only shell-1 neighbors and the connections among them. The residual size is therefore independent of \ensuremath{y} and \ensuremath{w}, and is reduced to \ensuremath{m_{kd}^{(1)}}. Equation~(\ref{9}) then reduces to
\begin{equation}\label{4+}
\epsilon_k
=
q
\sum_d
m_{kd}^{(1)}P(d\mid k)\epsilon_d .
\end{equation}
This equation coincides with the message-passing equation in
Refs.~\cite{Serrano2006,Serrano2006b,Serrano2006PRL}, where $m_{kd}^{(1)}$
counts the remaining edges after excluding the incoming edge and the edges
that form triangles with it.

At zeroth order, the subgraph contains only shell-1 neighbors, so the residual size is the degree of the reached node minus one, i.e., \ensuremath{d-1}. One obtains
\begin{equation}\label{4}
\epsilon_k
=
q
\sum_d
(d-1)P(d\mid k)\epsilon_d ,
\end{equation}
which is the locally tree-like message-passing equation.

Together, Eqs.~(\ref{4+}) and~(\ref{4}) show that the zeroth- and first-order equations are strict lower-order limits of the same generalized-edge framework, obtained by discarding the connectivity variables $y$ and $w$ that first appear at $n=2$; the corresponding derivation is given in Supplemental Material. The second-order formulation further retains the effective connectivity inside the generalized edge $E_{kd}^{(2)}$,
leading to the tractable higher-order message-passing equation, Eq.~(\ref{9}). This equation will be used below to analyze both synthetic and real networks.

\subsection{Validity Analysis of the Approximation}

To assess the approximation at the current order, we first characterize the
generalized-edge count distribution induced by the corresponding local
subgraphs, which provides a basic description of the size and heterogeneity
of the generalized-edge space. The validity of the approximation is then assessed through the closure
property of generalized edges, as quantified by GECC. The resulting value indicates whether the approximation at the current order is reliable.

\subsubsection{Statistics of Generalized Edges and the Induced Generalized Degree Distribution}

We therefore first characterize the generalized degree distribution, defined
as the distribution of the number of generalized edges rooted at each node.
Let $G$ denote the whole graph. For a node $u\in G$, let
$\Gamma_u^{(n)}$ denote the $n$th-order subgraph of node $u$. For each
node $v\neq u$ contained in $\Gamma_u^{(n)}$, the node pair $(u,v)$ labels an
$n$th-order generalized edge rooted at $u$, whose structure is given by the
intersection $\Gamma_u^{(n)}\cap\Gamma_v^{(n)}$. We collect the distinct labels of all $n$th-order generalized edges in $G$ as
\begin{equation}\label{ge_set}
\mathcal{R}_{G}^{(n)}
=
\left\{
(u,v):
u\in G,\ 
v\in \Gamma_u^{(n)},\
v\neq u
\right\}.
\end{equation}
The generalized degree of $u$ is therefore
\begin{equation}\label{ge_dist_1}
K_{u}^{(n)}
=
\left|\Gamma_u^{(n)}\right|-1 .
\end{equation}
The corresponding generalized degree distribution is defined as
\begin{equation}\label{ge_dist_2}
P_{\mathrm{GE}}^{(n)}(K)
=
\frac{1}{\left|G\right|}
\sum_{u\in G}
\mathbf{1}\!\left[
K_{u}^{(n)}=K
\right].
\end{equation}
Here, \(\mathbf{1}[\cdot]\) denotes the indicator function, which equals one if
the condition inside the brackets is satisfied and zero otherwise. For $n=1$, this distribution reduces to the ordinary degree distribution. For larger $n$, $P_{\mathrm{GE}}^{(n)}(K)$ characterizes how rapidly the subgraph expands with the retained order, thereby describing the size and heterogeneity of the generalized-edge space used in the closure.

\subsubsection{Generalized-edge closure coefficient}

The threshold predicted by the recursion in Eq.~(\ref{9}) is accurate
only when a generalized edge behaves as a single effective structure.
Concretely, consider a generalized edge induced by an initial node
pair. Because a message may enter this structure through any node pair
internal to it, each such internal pair can equally be used to
induce a generalized edge. Choosing a different internal node pair
therefore amounts to probing the same underlying structure from a
different direction of message flow. If every internal node pair selected within this generalized edge re-induces
the same generalized edge, then the structure identified does not
depend on the direction from which it is entered: the generalized
edge indeed behaves as a single large node, messages entering it from
any direction pass through the same structure, and the recursion
yields an accurate threshold prediction.

This need not hold. When an internal node pair induces a different
generalized edge, the structure identified depends on the direction
from which it is entered: the induced structure changes with the
choice of node pair used to probe it. This is illustrated by the
white node in Fig.~\ref{fig:ge_schematic_summary}(C): viewed from the
original node pair (left panel), the white node is absorbed into the
generalized edge and does not participate in subsequent message
aggregation; viewed from an internal pair within the same structure
(right panel), the same node instead falls into the residual
structure and re-enters the message-passing sum, introducing an
error in the predicted threshold. Because the network structure being
probed does not itself change, this inconsistency is not a property
of the white node itself, but a signature that the generalized edge
fails to provide a well-defined edge/residual partition. As a result, nodes that should have been absorbed into the generalized
edge are instead erroneously reintroduced into the residual structure,
introducing an error into the message-passing equation.

We refer to a generalized edge as having good generalized-edge closure when
most internal node pairs selected within the original generalized edge re-induce the same generalized edge. Conversely, when only a small fraction of
internal node pairs re-induce the same generalized edge, or when many of them
induce different generalized edges, the generalized edge is regarded as having
poor generalized-edge closure. Therefore, better generalized-edge closure
indicates that the corresponding message-passing equations are more likely to
yield accurate threshold predictions.

This notion differs from conventional measures of clustering or short-loop density, including clustering coefficients~\cite{Watts1998,Newman2003}, motif and cycle-based statistics~\cite{Milo2002}, and edge multiplicity~\cite{Serrano2006,Serrano2006b,Serrano2006PRL}. These measures either quantify triangle closure at the node, edge, or network level, or rely on predefined subgraph patterns or cycle structures, and thus are insufficient for probing the internal structure of the overlap between arbitrary subgraphs. This limitation motivates the introduction of a new structural statistic for generalized-edge closure.

To quantify this property, we introduce an order-dependent generalized-edge closure coefficient, \ensuremath{\mathrm{GECC}^{(n)}}. For each
\ensuremath{(u,v)\in\mathcal{R}_{G}^{(n)}}, let
\ensuremath{n_{uv}=|\Gamma_u^{(n)}\cap\Gamma_v^{(n)}|}; by construction, the
reference nodes \ensuremath{u} and \ensuremath{v} belong to
\ensuremath{\Gamma_u^{(n)}\cap\Gamma_v^{(n)}}, so \ensuremath{n_{uv}\geq 2}.
Then we consider all unordered node pairs
\ensuremath{(x,y)} within \ensuremath{\Gamma_u^{(n)}\cap\Gamma_v^{(n)}} and ask whether the generalized edge labeled by \ensuremath{(x,y)} coincides with that labeled by
\ensuremath{(u,v)}. The local closure-inconsistency score is
\begin{equation}\label{17}
\zeta_{uv}^{(n)}
=
1-
\frac{
\sum_{\substack{x,y\in\Gamma_u^{(n)}\cap\Gamma_v^{(n)}\\ x<y}}
\mathbf{1}
\left[
\Gamma_x^{(n)}
\cap
\Gamma_y^{(n)}
=
\Gamma_u^{(n)}
\cap
\Gamma_v^{(n)}
\right]
}{
\binom{n_{uv}}{2}
},
\end{equation}
which gives the fraction of internal node pairs that fail to recover the same generalized edge; the condition \ensuremath{x<y} simply avoids double counting unordered node pairs. When \ensuremath{n_{uv}=2}, the only internal pair is the reference pair \ensuremath{(u,v)} itself, hence
\ensuremath{\zeta_{uv}^{(n)}=0}, in accordance with the fact that such a generalized edge degenerates into an
ordinary edge with perfect closure. Averaging over all generalized edges yields 
\begin{equation}\label{19}
\mathrm{GECC}^{(n)}
=
\left\langle
\zeta_{uv}^{(n)}
\right\rangle_{\mathcal{R}_{G}^{(n)}} .
\end{equation}
The detailed computational procedure for evaluating
\ensuremath{\mathrm{GECC}^{(n)}} from network data is provided in the
Supplemental Material.

Small values of \ensuremath{\mathrm{GECC}^{(n)}} indicate good generalized-edge closure: generalized edges are consistently recovered from their internal node pairs, yielding a reliable $n$th-order approximation. Large values indicate poor closure: the recovered generalized
edges depend strongly on the choice of internal reference pair, making the
$n$th-order approximation less reliable. Below, we use
\ensuremath{\mathrm{GECC}^{(1)}} and \ensuremath{\mathrm{GECC}^{(2)}} to assess
the closure of the corresponding approximations on synthetic and real networks.

\subsection{Application on synthetic and real networks}

\subsubsection{Threshold prediction diagnostics}

We evaluate the predictive accuracy of approximations of different orders
for site percolation. For each network, we compare the simulation-based
threshold \ensuremath{q_c^{\mathrm{sim}}} with three theoretical estimates:
the locally tree-like prediction \ensuremath{q_c^{\mathrm{tree}}}, the
first-order threshold prediction \ensuremath{q_c^{(1)}}, and the
second-order threshold prediction \ensuremath{q_c^{(2)}}. The numerical
procedures used to estimate these thresholds are described in Methods. We
then compute \ensuremath{\mathrm{GECC}^{(1)}} and \ensuremath{\mathrm{GECC}^{(2)}}
and relate them to the gaps between the predicted and simulated
thresholds.

\begin{figure*}[t]
    \centering
    \includegraphics[width=0.95\textwidth]{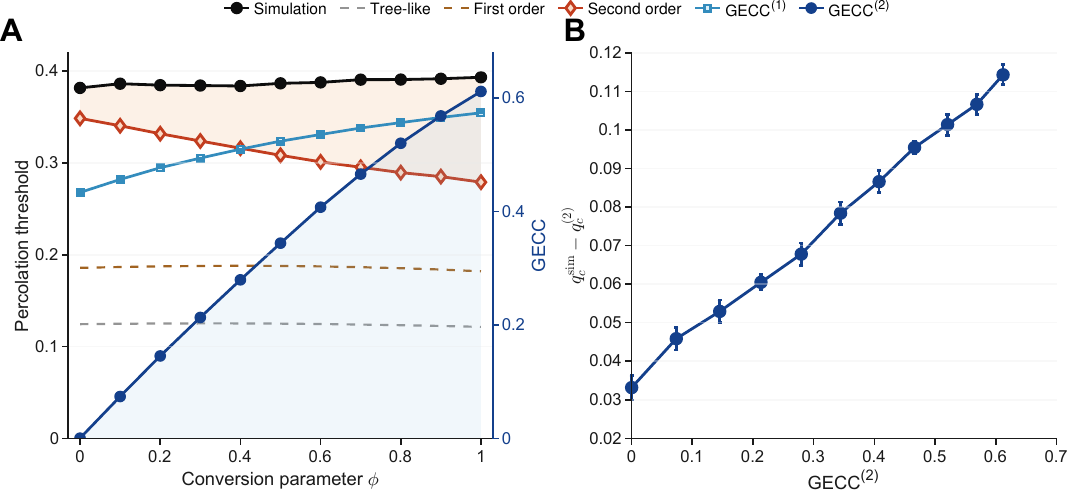}
  \caption{
\textbf{Closure departure drives finite-order threshold failure in
synthetic networks.}
\textbf{(A)} Simulated percolation thresholds are compared with locally
tree-like, first-order, and second-order threshold predictions as the
conversion parameter $\phi$ is varied. Threshold quantities are
plotted on the left axis, and $\mathrm{GECC}^{(1)}$ and
$\mathrm{GECC}^{(2)}$ are plotted on the right axis. Increasing $\phi$
drives the generalized edges from the near-closed regime (small
$\phi$) into the closure-departure regime (large $\phi$). The orange
shaded region marks the residual gap between $q_c^{(2)}$ and
$q_c^{\mathrm{sim}}$. The pale-blue shading provides a visual guide
for $\mathrm{GECC}^{(2)}$. Error bars are included for
$q_c^{\mathrm{sim}}$ and both GECC curves, but they are smaller than,
or comparable to, the symbol size.
\textbf{(B)} The residual threshold gap $q_c^{\mathrm{sim}}-q_c^{(2)}$ is
plotted directly against $\mathrm{GECC}^{(2)}$. The two
quantities trace out an approximately linear relationship across the
full range of $\phi$, directly showing that the loss of second-order
closure quantitatively accounts for the degradation of second-order
threshold accuracy, rather than merely being correlated with it.
}
    \label{fig:threshold_gecc_synthetic}
\end{figure*}

\subsubsection{Validation using synthetic networks}

We first test whether \ensuremath{\mathrm{GECC}^{(n)}} tracks the
reliability of the approximations on synthetic networks with controlled
generalized-edge closure. The networks are generated by a two-step unit
replacement procedure on a random backbone. Each backbone edge is first
replaced by multiple type-I local units, and a fraction of these units
is then converted into type-II units. This construction produces tunable
loop-mediated dependencies while keeping the global backbone fixed; the
detailed replacement procedure is described in Methods. The conversion
parameter \ensuremath{\phi} controls the fraction of type-I units
converted into type-II units, thereby providing a controlled way to tune
second-order closure quality while limiting changes in first-order
closure.

As the conversion parameter \ensuremath{\phi} increases, the tree-like and
first-order predictions remain well below \ensuremath{q_c^{\mathrm{sim}}}
throughout the sweep, confirming that these low-order approximations are
insensitive to the loop-mediated dependencies introduced by the unit
replacement. At small \ensuremath{\phi}, the second-order generalized edges remain
well closed at second order, and \ensuremath{q_c^{(2)}} follows the
simulation closely, with only a small residual gap attributable to
loops already present in the fixed backbone. As
\ensuremath{\phi} increases, internal re-induction inconsistencies
accumulate, \ensuremath{\mathrm{GECC}^{(2)}} rises monotonically, and the
residual second-order threshold error grows in step, while
\ensuremath{\mathrm{GECC}^{(1)}} remains at a moderate-to-large, weakly
varying level. Because \ensuremath{\phi} is the only parameter varied and
acts specifically on second-order local structure, this co-response
isolates closure loss, rather than a generic increase in topological
complexity, as the mechanism driving finite-order failure. The sweep
thereby constitutes a controlled mechanism test linking closure loss to
finite-order prediction error; we next examine whether the same relation
persists in uncontrolled, naturally occurring network structures.

\subsubsection{Real-world network analysis}

\begin{figure}[t]
    \centering
    \includegraphics[width=\columnwidth]{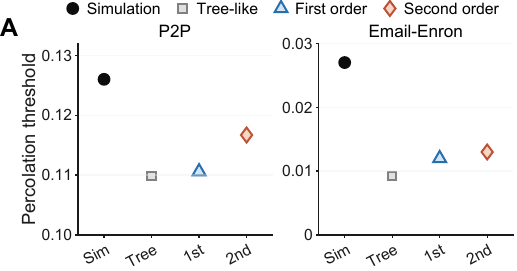}

    \vspace{0.6em}

    \includegraphics[width=\columnwidth]{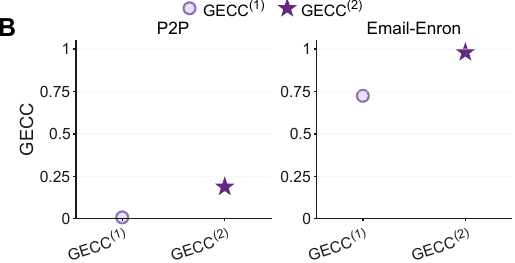}

    \caption{
    \textbf{GECC as an a priori diagnostic of finite-order threshold
    reliability in representative real networks.}
    \textbf{(A)} Simulation thresholds and theoretical predictions for P2P and
    Email-Enron. The thresholds are plotted on ordinary linear axes, allowing
    direct comparison between \ensuremath{q_c^{\mathrm{sim}}},
    \ensuremath{q_c^{\mathrm{tree}}}, \ensuremath{q_c^{(1)}}, and
    \ensuremath{q_c^{(2)}}.
    \textbf{(B)} Corresponding generalized-edge closure coefficients
    \ensuremath{\mathrm{GECC}^{(1)}} and
    \ensuremath{\mathrm{GECC}^{(2)}} for the same two networks. Larger GECC
    values indicate poorer generalized-edge closure. P2P has small GECC
    values and a correspondingly small residual threshold error, whereas
    Email-Enron has \ensuremath{\mathrm{GECC}^{(2)}} close to one and a large
    residual error, illustrating the two structurally distinct regimes
    identified across all four networks in Table~\ref{tab:real_networks}.
    }
    \label{fig:real_networks_threshold_gecc}
\end{figure}

Having established the closure-departure mechanism on controlled synthetic
benchmarks, we next apply the same threshold-diagnostic protocol to
real-world networks from the Stanford Large Network Dataset
Collection~\cite{snapnets}, to test whether the same mechanism, and the
same order-adequacy classification, extends to naturally occurring, uncontrolled network structures. For visual clarity,
Fig.~\ref{fig:real_networks_threshold_gecc} shows two representative
networks, P2P and Email-Enron, while the full results are summarized in
Table~\ref{tab:real_networks}.

\begin{table*}[t]
\centering
\caption{Percolation thresholds and GECC values for real networks.}
\label{tab:real_networks}
\begin{ruledtabular}
\begin{tabular}{lcccccc}
Network
& $q_c^{\mathrm{sim}}$
& $q_c^{\mathrm{tree}}$
& $q_c^{(1)}$
& $q_c^{(2)}$
& $\mathrm{GECC}^{(1)}$
& $\mathrm{GECC}^{(2)}$ \\
Facebook
& 0.48
& 0.0073
& 0.028
& 0.040
& 0.98
& 0.99 \\
GitHub
& 0.27
& 0.0076
& 0.0094
& 0.010
& 0.65
& 1.00 \\
P2P
& 0.13
& 0.11
& 0.11
& 0.12
& 0.0076
& 0.19 \\
Email-Enron
& 0.027
& 0.0092
& 0.012
& 0.013
& 0.72
& 0.98 \\
\end{tabular}
\end{ruledtabular}
\end{table*}

The real networks fall into two structurally distinct regimes. P2P has
comparatively small GECC values, especially \ensuremath{\mathrm{GECC}^{(1)}},
and is predicted accurately at low order, consistent with the synthetic
near-closed regime identified above. Facebook, GitHub, and Email-Enron
instead have \ensuremath{\mathrm{GECC}^{(2)}} values close to one and retain
large residual threshold errors, indicating that their generalized edges
are poorly closed at second order. Across all four networks,
\ensuremath{q_c^{(2)}} is generally closer to \ensuremath{q_c^{\mathrm{sim}}}
than \ensuremath{q_c^{(1)}}, but for Facebook, GitHub, and Email-Enron this
improvement does not translate into an accurate second-order prediction,
exactly as anticipated by their \ensuremath{\mathrm{GECC}^{(2)}} values.
Notably, because \ensuremath{\mathrm{GECC}^{(2)}} for these three networks is
already close to its upper bound, it can only signal that second-order
closure is inadequate without further resolving how severely, or in what
structural respect, the closure fails; distinguishing among these networks
and assessing whether third- or higher-order truncation would meaningfully
reduce the residual error therefore requires evaluating
\ensuremath{\mathrm{GECC}^{(n)}} for \ensuremath{n\geq 3}.

This agreement between closure status and residual error supports two
conclusions. First, it is not merely correlational: because
\ensuremath{\mathrm{GECC}^{(n)}} is computed from local structure alone,
without reference to \ensuremath{q_c^{\mathrm{sim}}}, its value correctly
anticipates, prior to any percolation simulation, whether a given
truncation order will be adequate. In this sense,  the apparent failure of
the second-order approximation on three of the four networks is not a
counterexample to the theory but a confirmation of it: GECC identifies in
advance exactly which networks are poorly closed at second order. Second,
the same order-adequacy classification that emerged from the controlled synthetic sweep discussed above --- where closure loss, rather than generic
topological complexity, was shown to drive finite-order failure --- carries
over unchanged to naturally occurring, uncontrolled network structures.
Together, these results elevate \ensuremath{\mathrm{GECC}^{(n)}} from a
descriptive correlate of threshold prediction error into an \emph{a priori}
validity certificate for finite-order message passing, usable to decide,
from structure alone, when a given order suffices and when the residual
loop-induced correlations require a higher-order treatment.

\section{Discussion}

We have developed an ensemble-level loopy message-passing framework for
percolation on highly clustered random graphs. Building on previous work, the
framework extends progressively refined approximations to random graph
ensembles, thereby generalizing the study of loop-induced correlations from
predefined motif structures to arbitrarily structured local neighborhoods. In
doing so, the framework formulates the problem at the ensemble level and
therefore does not require the full adjacency matrix. It closes the resulting
equations using higher-order local network features extracted from ensemble
statistics, extending beyond degree distributions and degree correlations to
capture loop-induced dependencies in highly clustered random graphs. In this
formulation, these dependencies are represented through generalized edges and
the overlaps between them, allowing local clustering effects to be incorporated
directly into the ensemble-level equations.

A central and perhaps counterintuitive finding is that the accuracy of a
low-order approximation is not controlled solely by the density of short
loops. It also depends on whether the internal structure of a generalized
edge is consistent at the current order: reselecting any node pair within a
generalized edge should reproduce the same generalized-edge structure.
Failure of this consistency indicates that the generalized edge does not
form a stable closure unit at the retained order, which can bias the
predicted threshold. The generalized-edge closure coefficient,
\ensuremath{\mathrm{GECC}}, provides a diagnostic for this effect. In
synthetic networks with controlled local motifs, \ensuremath{\mathrm{GECC}^{(1)}}
is large whereas \ensuremath{\mathrm{GECC}^{(2)}} is small. Consistently, the
second-order approximation accurately reproduces the simulated percolation
threshold, whereas the tree-like and first-order approximations do not. Real
networks show the same pattern: networks with larger second-order closure
deficits, such as Facebook, GitHub, and Email-Enron, exhibit larger residual
threshold errors, whereas networks with smaller deficits, such as P2P, are
predicted more accurately. Thus, improving loopy message passing requires
not only adding more short-loop statistics, but also ensuring the internal
consistency of generalized edges at the retained order.

This framework opens three natural directions for future work: analyzing
critical exponents and scaling behavior, systematically extending the
closure to higher orders, and generalizing the closure diagnostic beyond
percolation. Since the theory yields ensemble-level self-consistent
equations for percolation on highly clustered random graphs, it provides a
basis for examining how loop-induced correlations affect not only the
percolation threshold but also the critical regime near the transition. A
second direction is to improve threshold-prediction accuracy while keeping
the increase in approximation order as small as possible. Two limitations
of the present approach motivate this direction. First, our validation has
been carried out only up to second order; how the computational and
identification cost of \ensuremath{\mathrm{GECC}} scales with order has not
been systematically assessed. Second, \ensuremath{\mathrm{GECC}} itself is
a diagnostic with an inherent blind spot: a small value of
\ensuremath{\mathrm{GECC}^{(n)}} certifies that order-\ensuremath{n} closure
is adequate, but it cannot on its own rule out the presence of longer-range
correlations that only become visible at order \ensuremath{n+1} or beyond.
Determining the appropriate closure order therefore requires computing
\ensuremath{\mathrm{GECC}^{(n)}} progressively until it decays, rather than
relying on any single low-order value. This requires first discerning how
strongly longer loops influence message passing, so as to properly balance
the accuracy loss incurred by neglecting such loops against the cost of
identifying the corresponding generalized edges and computing
\ensuremath{\mathrm{GECC}} at increasing order. Determining the closure
order at which the improvement in threshold prediction or critical scaling
becomes marginal is therefore an important problem. \ensuremath{\mathrm{GECC}}
may provide a practical diagnostic for this purpose, indicating when
higher-order corrections are necessary and when the current level of
closure is sufficient. A third direction is to extend this closure
diagnostic beyond percolation to other network processes that are similarly
sensitive to loop-induced correlations, such as cascading failures,
epidemic spreading, and opinion dynamics, since the underlying notion of a
generalized edge and its internal consistency does not depend on the specific dynamical process under study.

\section{Methods}

\subsection{First- and second-order threshold predictions}

For a given network, self-loops and duplicate edges were first removed, and
nodes were relabeled by consecutive integers. For each node $u$,
nodes $v$ were sampled from the second-order subgraph $\Gamma_u^{(2)}$,
which was used to estimate the quantities entering the second-order
message-passing matrix. The second-order subgraph $\Gamma_v^{(2)}$ was then
extracted, so that the generalized edge labeled by the pair $(u,v)$ could be
explicitly determined. Degree-stratified sampling was used with total budget
$N_{\rm sample}=10^7$: if $\mathcal{D}$ is the set of degrees present in the
network and $N_k$ is the number of nodes with degree $k$, each node of degree
$k$ was assigned
$\lceil N_{\rm sample}/(|\mathcal{D}|N_k)\rceil$ samples; if the number of
available nodes $v$ in $\Gamma_u^{(2)}$ was smaller than this assigned budget,
all such nodes were used.

For each sampled ordered generalized edge $(u,v)$, we recorded
$(k,d,y,w,m)$, with $k=\deg u$ and $d=\deg v$, where $y$ and $w$ denote the
second-order overlap variables characterizing the internal structure of the
generalized edge, and $m$ denotes its multiplicity. These records were used
to estimate $P(d,y,w\mid k)$ and $m_{kdyw}^{(2)}$. The second-order matrix was
constructed as
\begin{equation}
M_{kd}^{(2)}(q)
=
q\sum_{y,w}
m_{kdyw}^{(2)}
g(q,y,w)
P(d,y,w\mid k),
\end{equation}
and the second-order threshold prediction $q_c^{(2)}$ was obtained from the
condition $\rho(M^{(2)}(q_c^{(2)}))=1$, where $\rho(\cdot)$ denotes the
spectral radius of a matrix.

The first-order threshold prediction $q_c^{(1)}$ was obtained using the same
procedure restricted to the order-one subgraph $\Gamma_u^{(1)}$. In this case,
ordered generalized edges were sampled from $\Gamma_u^{(1)}\setminus\{u\}$,
and the second-order structural variables $y$ and $w$ were not recorded.

Details are provided in Supplemental Material.

\subsection{Synthetic network construction}

Synthetic networks were constructed by a two-step unit replacement procedure
starting from a simple random $d_0$-regular backbone with $N_0$ nodes
($N_0=1000$, $d_0=4$ in this study). In the first step, each backbone edge
$e=(a,b)$ was removed and replaced by $m_e$ type-I local units, where $m_e$
was drawn uniformly from a fixed integer range ($m_{\min}=3$, $m_{\max}=8$ in
this study). Each type-I unit introduced two new nodes, $c$ and $d$, and the
five edges $(a,c)$, $(a,d)$, $(b,c)$, $(b,d)$, and $(c,d)$. Thus, the two
backbone nodes $a$ and $b$ were no longer connected by the original backbone
edge, but were connected through $m_e$ dense local units.

In the second step, a fraction of the type-I units was converted into a second
unit type. The conversion parameter, denoted by $\phi$, specifies the fraction
of type-I units selected uniformly at random for this conversion. For each
selected unit, the original two new nodes were identified as $2=c$ and $5=d$,
and the backbone nodes were identified as $1=a$ and $6=b$. Two additional
nodes, $3$ and $4$, were then added. The five type-I edges were removed and
replaced by the nine type-II edges $(1,2)$, $(2,3)$, $(3,6)$, $(1,4)$,
$(2,5)$, $(4,5)$, $(5,6)$, $(2,4)$, and $(2,6)$. Therefore, increasing
$\phi$ increases the fraction of local connectivity patterns with the type-II
closure pattern while preserving the same type of global backbone.

\section*{Supplemental Material}

See the Supplemental Material for detailed definitions, derivations,
algorithmic implementation, and additional numerical results.

\section*{Data availability}

All MATLAB code and scripts required to reproduce the computational analyses
in this study, including the scripts used to generate the synthetic networks,
together with the processed real network edge-list data used in the analyses,
will be made publicly available upon publication in the GitHub repository
\url{https://github.com/qwerwerv/Percolation-and-GECC-code} and in the Zenodo
archive \url{https://doi.org/10.5281/zenodo.21390642}. The original real-world
network datasets analyzed in this study were obtained from the Stanford Large
Network Dataset Collection~\cite{snapnets}.

\begin{acknowledgments}
This work was supported by the Science Challenge Project under Grant
No. TZ2025017.
\end{acknowledgments}

\nocite{*}

\bibliography{apssamp}% Produces the bibliography via BibTeX.

\end{document}